\documentclass[sigconf]{acmart}
\usepackage{booktabs} % For formal tables
\usepackage[symbol]{footmisc}
\usepackage[ruled]{algorithm2e} % For algorithms

\SetAlFnt{\small}
\SetAlCapFnt{\small}
\SetAlCapNameFnt{\small}
\SetAlCapHSkip{0pt}
\IncMargin{-\parindent}

\setcopyright{rightsretained}
\newfont{\mycrnotice}{ptmr8t at 7pt}
\newfont{\myconfname}{ptmri8t at 7pt}

\begin{document}

\acmConference[LBRS@RecSys '18]{ACM Woodstock conference}{Vancouver, Canada}{October, 2018}
\acmYear{2018}
\copyrightyear{2018}

% \acmPrice{15.00}

% \acmSubmissionID{123-A12-B3}

\BeforeBeginEnvironment{figure}{\vskip-1.5ex}
\AfterEndEnvironment{figure}{\vskip-2.5ex}
% \BeforeBeginEnvironment{table}{\vskip-2ex}
% \AfterEndEnvironment{table}{\vskip-5ex}

\title{Content-based feature exploration for transparent music recommendation using self-attentive genre classification}

\author{Seungjin Lee}
\affiliation{%
  \institution{Seoul National University}
%   \streetaddress{Seoul, Korea}
%   \state{Seoul}
%   \city{Korea}
}
\email{joshua77@snu.ac.kr}

\author{Juheon Lee}
\affiliation{%
  \institution{Seoul National University}
% %   \streetaddress{Seoul, Korea}
%   \state{Seoul}
%   \city{Korea}
% %   \postcode{43017-6221}
}
\email{juheon2@snu.ac.kr}

\author{Kyogu Lee}
\affiliation{%
  \institution{Seoul National University}
% %   \streetaddress{Seoul, Korea}
%   \state{Seoul}
%   \city{Korea}
% %   \postcode{43017-6221}
}
\email{kglee@snu.ac.kr}

% \affiliation{%

% \renewcommand{\shortauthors}{B. Trovato et al.}

\begin{abstract}
Interpretation of retrieved results is an important issue in music recommender systems, particularly from a user perspective. In this study, we investigate the methods for providing interpretability of content features using self-attention. We extract lyric features with the self-attentive genre classification model trained on 140,000 tracks of lyrics. Likewise, we extract acoustic features using the acoustic model with self-attention trained on 120,000 tracks of acoustic signals. The experimental results show that the proposed methods provide the characteristics that are interpretable in terms of both lyrical and musical contents. We demonstrate this by visualizing the attention weights, and by presenting the most similar songs found using lyric or audio features. 
% , and comparable result with existing methods in acoustic signals as 64\%.
\end{abstract}




\maketitle

\section{Introduction}

In terms of real world applications, it is important to provide explainable recommendation. Explanations may show how the system works or help users make an informed choice \cite{two}. 
% Users usually want to explore recommendation results while using recommendation services.
Providing explanations usually build user's trust in recommender system and it's results. In order to extract features, many existing systems in content-based music recommendation use music classification which is an well-researched task in Music Information Retrieval. However, extracted features from these methods only provide high-level explanation (e.g., similar artist or style). 

The focus of this paper is a methods for low-level explanation of recommendation results (e.g., certain words of lyric reflecting genre, important part of acoustic signal reflecting genre). In order to extract explainable content features self-attention architecture is used with genre classification. \cite{five} extract attention weights of lyric using hierarchical attention networks which hierarchically attend segments, lines and words. However, it is more reasonable to provide entire song-level explanation to users. \cite{six} propose a self-attention network for sentence embedding, visualizing important words or phases using attention weights. Inspiring this work, we formulate entire song-level self-attentive genre classification task using two different dataset (Lyric and Audio) respectively.

\section{DATASET}

Since there is no large lyric dataset which is publicly available, we use  dataset\footnote{\label{myfootnote}https://www.kaggle.com/gyani95/380000-lyrics-from-metrolyrics/home} collected from MetroLyrics which has lyrics database featuring 1,000,000+ song lyrics from 20,000 artists. This consist of single label with 10 genre classes. We use subset of Top MAGD dataset\footnote{\label{myfootnote}http://www.ifs.tuwien.ac.at/mir/msd/TopMAGD.html} as Audio dataset. This also consist of single label with 13 genre classes.
We reject the lyrics whose length is less than 70 words and the audio less than 30 seconds.  

\section{self-attention architecture}
The entire framework of the model is illustrated in Figure 1. 
% We use bidirectional Long Short-Term Memory (BiLSTM) with self attention applied to output of BiLSTM. 
Unlike previous research, we use pre-trained embedding rather than index of word dictionary as input to the model. The song lyrics are split into words and they are represented as a 500 by 128 matrix using pre-trained Word2Vec model \cite{word2vec}. To change the task into an entire song-level task, we set the word length to 500 with zero padding for lyrics less than 500 words. The 30 seconds of audio signals are converted to mel-spectrograms and they are represented as a 30 (length of signals sliced by 1 seconds) by 128 matrix using pre-trained Vggish model \cite{vggish}. 

\begin{figure}[!h]
  \centering
  \includegraphics[height=4.9cm, width=8.5cm]{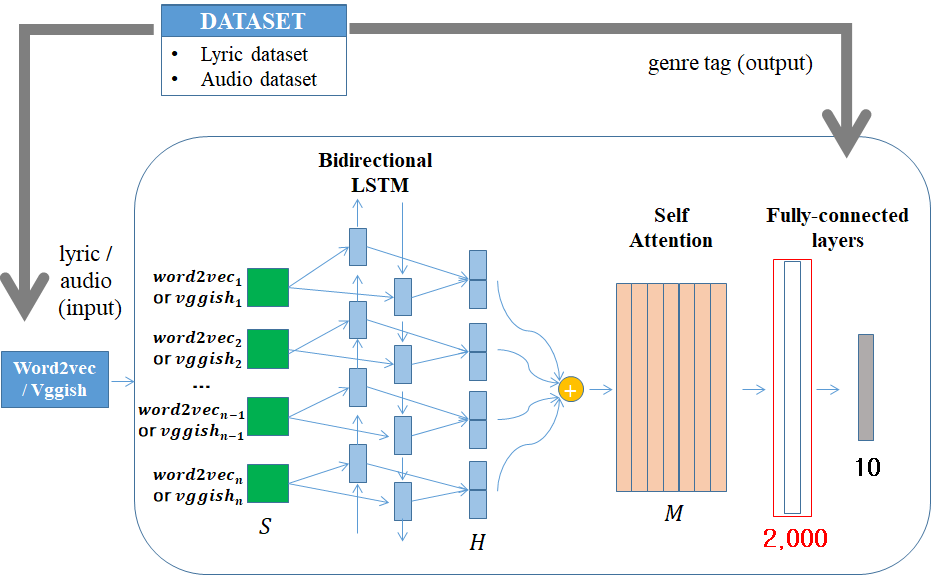}
  \caption{Model structure for self-attentive genre classification}
\end{figure}
Given outputs of BiLSTM \(H\), the attention can be formulated as
\begin{gather}
  A = softmax(W_{s2}\cdot tanh(W_{s1}\cdot H^T)) \label{eq1} \\
  M = A \cdot H \label{eq2} 
\end{gather}
\begin{equation}
       A_i = 
        \begin{cases}
            A_i & \text{if $A_i > 0.15 \cdot max(A)$} \\
            0 & \text{otherwise}
        \end{cases}
\end{equation}
\(W_{s1}\) and \(W_{s2}\) are learned by the model after random initialization and the shape of these parameter matrix is hyperparameter except for input length.  
Matrix \(M\) is flattened for genre classification and attention matrix \(A\) is used for visualization of words or sliced audio signals. After training, outputs include content embedding with 2,000 dimensions and attention weights. In order to visualize attention weights, we set condition as in (3). 

\begin{figure}[!h]
  \centering
  \includegraphics[height=4cm, width=7.5cm]{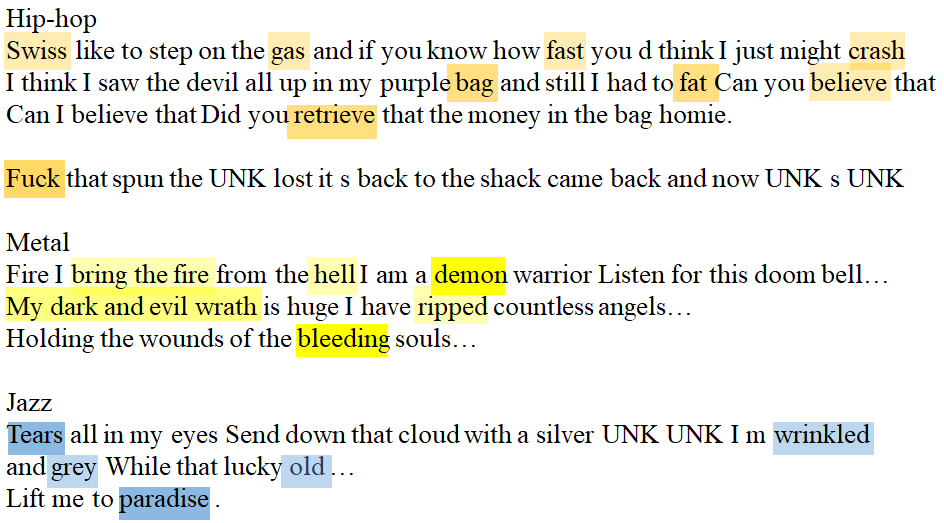}
  \caption{Visualization of attention weights of lyrics in hip-hop, metal and jazz genre}
\end{figure}

\begin{figure}[!h]
  \centering
  \includegraphics[height=2.5cm, width=8.5cm]{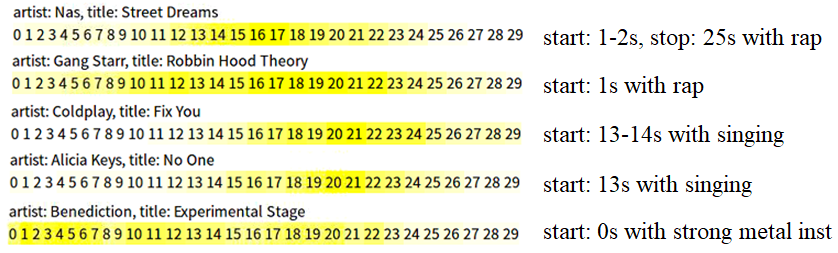}
  \caption{Visualization of attention weights of 30 seconds of acoustic signals}
\end{figure}

\begin{figure}[!ht]
  \centering
  \includegraphics[height=7cm, width=8cm]{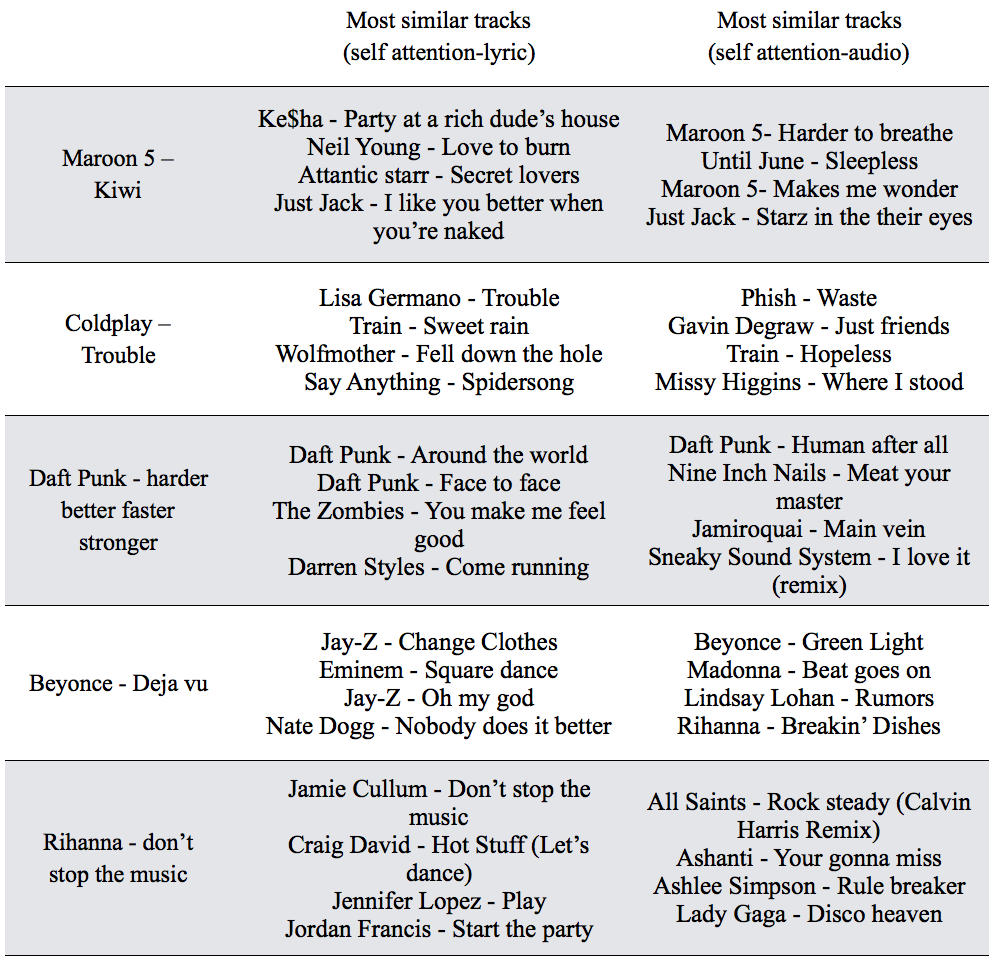}
  \caption{Result of similar song retrieval using Lyric (left) and Audio (right) dataset}
\end{figure}

\section{Experimental results}
As a result of visualizing attention weights of lyrics, there is a tendency to focus on words that have an important role in genre classification. Figure 2 shows the visualization of attention weights of lyrics in hip-hop genres. In this case, the model focus on strong and aggressive words such as slang. The model also attend to "rhyme" which is interesting features of hip-hop music. In addition, the metal genre tend to concentrate on aggressive but religious words related to good and evil. Visualization of the attention weights of acoustic signals show a tendency to focus on the beginning of vocal. Figure 3 shows the visualization of attention weights of acoustic signals. in the case of "fix you" of Coldplay, singing start from 13 to 14 seconds. In the case of song of Benediction which has strong metal sound from the beginning, Figure 3 shows that there is a strong attention from the beginning. This may be related to the fact that the pre-trained dataset for Vggish model include diverse speech and instrument data.
We explore similar song with query using extracted content features. Figure 4 shows the results of top 4 similar songs with query. Retrieved similar songs from lyrics include cover songs, an important word of lyrics and mood of lyrics. An important word in "trouble" of Coldplay is a "spider", and "spidersong" is retrieved as similar song. Even though the title is not trained, songs with the same title are retrieved such as "trouble" of Lisa Germano. Retrieved similar songs from acoustic signals include similar genre, similar artists and similar vocal timbre. "Don't stop the music" of Rihanna is exciting dance music with female vocal sound, and the dance music with female vocal sound is retrieved as similar songs of Rihanna's music.

\section{Conclusions}
In this research, we extract content features using self-attentive genre classification, and explore extracted features and attention weights. We indicate acoustic signals are as good features as lyrics in terms of interpretation, and both of them reflect the details that they do not reflect each other. We will further investigate the performance with same dataset of lyric and audio, and compare two modalities in terms of recommendation. Furthermore, we will examine the feasibility of explainable recommendation in real-world applications.

\section{Acknowledgements}
This work was supported by Kakao and Kakao Brain Corp.

\bibliographystyle{ACM-Reference-Format}
\bibliography{mybib}

\end{document}